\documentclass[twocolumn,showpacs,superscriptaddress,prb]{revtex4}
\usepackage{amsbsy,amssymb,amsmath,bm} \usepackage{graphicx}
\usepackage{ulem} 
\usepackage{epsfig}
\usepackage[usenames]{color}  %

\newcommand{\bk}{\mathbf{k}}

 \newcommand{\cF}{\mathcal{F}}
 
\newcommand{\cT}{\mathcal{T}}

\begin{document}

\title{Heat capacity of a thin membrane at very low temperature}

\author{O. V. Fefelov}
\email{olegfe@fys.uio.no}
\affiliation{Department of Physics, University of Oslo, PO Box
1048 Blindern, 0316 Oslo, Norway}
\author{J. Bergli}
\affiliation{Department of Physics, University of Oslo, PO Box
1048 Blindern, 0316 Oslo, Norway}
\author{Y. M. Galperin}
\affiliation{Department of Physics and Center
for Advanced Materials and Nanotechnology, University of Oslo, PO Box
1048 Blindern, 0316 Oslo, Norway}
\affiliation{A. F. Ioffe Physico-Technical Institute of Russian
  Academy of Sciences, 194021 St. Petersburg, Russia}
\affiliation{Argonne National Laboratory, 9700 S. Cass Av., Argonne,
  IL 60439, USA}

\begin{abstract}
We calculate the dependence of heat capacity of a free standing thin
membrane on its thickness and temperature. A remarkable fact is
that for a given temperature there exists a minimum  in the
dependence of the heat  capacity on the thickness. The ratio of
the heat capacity to its minimal value 
for a given temperature is a universal function of the ratio of the
thickness to its value corresponding to the minimum. The minimal
value of the heat capacitance for given temperature is proportional to
the temperature squared. Our analysis can be used, in particular,  for
optimizing support membranes for microbolometers.
\end{abstract}
\maketitle

\paragraph{Introduction:}
Thin free standing membranes are extensively used for sensing and
detecting, in particular, for mounting of
microbolometers.\cite{paper2} Thermal and heat transport properties of
such membranes are very important for sensitivity of such bolometers
and their time response. At low temperatures the wavelength of
thermal phonons responsible for the heat capacity and conductance can
exceed the membrane thickness, $b$. In this case the vibrational modes
significantly differ from those in bulk materials.\cite{paper1}
In particular, the lowest vibrational mode has a {\em quadratic}
rather than linear 
dispersion law.\cite{book1}

The contributions of these modes to the low-frequency density of
vibrational states {\em increase} with decrease of the membrane
thickness. As a result, the low-temperature heat capacity also {\em
increases}. As the thickness increases, the heat capacity is
determined by higher modes having an essentially linear dispersion
law. Consequently, the heat capacity crosses over to that of a bulk
material. As a result, the thickness dependence of the heat capacity
of thin membranes is {\em nonmonotonous} having a sharp minimum at
some optimal, temperature dependent thickness. This minimum is
similar to the minimum in thickness dependence of the ballistic
heat transfer (power radiation) predicted in
Ref.~\onlinecite{paper3}. However, the 
concrete values of the optimal thickness for the heat capacity and
the ballistic heat transfer are different. Consequently, a proper
choice of the membrane thickness can be used for optimizing of the
time response of microbolometers mounted on thin free standing
membranes. The present paper is aimed at the theory of low temperature
heat capacity of thin free standing membranes.

\paragraph{Vibrational spectrum:}
The vibrational modes of a thin membrane are superpositions of
bulk longitudinal and transverse modes, their relative weights being
determined by boundary conditions -- both normal and tangential
stresses should vanish. The eigenmodes are classified as
 symmetric (SM) and antisymmetric (AM). 
Both are superpositions of the longitudinal and transverse bulk modes
with wave vectors $\bk^l$ and 
$\bk^t$, respectively. The relations between $\bk^l$ and $\bk^t$  are
\cite{book1}
\begin{eqnarray}\label{disp_rels}
\frac{\tan (b k^t_{\perp}/2)}{\tan ( b k^l_{\perp}/2)}& =& - \frac{4 k
^l_{\perp}
k^t_{\perp}k^2_{\parallel}}{[ (k^t)_{\perp}^2-k^2_{\parallel}]^2} \, , \\
\frac{\tan( b k^l_{\perp}/2)}{\tan (b k^t_{\perp}/2)}& =& - \frac{4 k^l_{\perp}
k^t_{\perp}k^2_{\parallel}}{[(k^t_{\perp})^2-k^2_{\parallel}]^2}
\label{disp_rela} 
\end{eqnarray}
for SM and AM, respectively. Here 
$k_\perp$ and $k_\parallel$ denote
perpendicular and parallel components of the wave vectors 
with  respect to the membrane.
Inserting the dispersion laws of the bulk modes,
  $k^l=\omega/c_l$ and $k^t=\omega/c_t$ where $c_l$ and $c_t$ are
  speeds of transversal and longitudinal sound, into Eqs.~(\ref{disp_rels})
 and (\ref{disp_rela}) one obtains 
transcendental equations 
for the dispersion laws, $\omega_{s ,n} (k_{\parallel})$,
of different vibrational branches. Here the subscript $\sigma$ stands
for the branch type while $n$ stands for its number.

In addition to AM and SM there
exists a horizontal shear mode (HS), which is a transversal wave with
both displacement and wave vector parallel to the plane of the
membrane. The HS mode dispersion law is\cite{book1} 
\begin{equation}\label{disp_relh}
\omega_{\text{HS},n}=c_t\sqrt{(n \pi/b )^2 + k^2_\parallel}
\end{equation}
where $n$ is an integer number. SM, AM and HS modes are the only
vibrations that can exist in a membrane. The modes possess a very
important property: as follows from Eqs.~(\ref{disp_rels}),
(\ref{disp_rela}), and (\ref{disp_relh}), the frequencies of all modes
scale as\cite{paper2} 
\begin{equation}
  \label{eq:scal}
  \omega_{s,n}=2c_t b^{-1}w_{s,n}(b k_\parallel)\, .
\end{equation}

Figure 1 shows the six lowest branches. 
Only the three lowest branches are gapless, consequently only they
contribute to the heat capacity of a membrane 
at $k_B T \ll \hbar \omega_{\text{HS},1}(0)$.
The lowest HS and SM are linear at $k \ll b^{-1}$ 
while the AM dispersion law can be approximated at small
  $k_\parallel$ as\cite{paper1}
\begin{equation} \label{ahlowest}
\omega_{\text{AM},0}(k_{\parallel}) = \frac{\hbar}{2m^{\ast}}
k_{\parallel}^2\, , \quad m^* \equiv \frac{\hbar  c_l \sqrt{3}}{2c_t  b
  \sqrt{c_l^2 - c_t^2} 
  } \, .
\end{equation}
The lowest AM branch is a flexural wave and the dispersion law of
this branch can be obtained from the 2D analog of Bernoulli-Euler
theory.\cite{book2} We will reproduce the derivation here since we
 will need it for analysis of the applicability range of our
 theory. Choosing the $z$-axis perpendicular to the membrane we
 have:\cite{book2} 
\begin{equation}\label{landau}
\rho \frac{\partial^2 u}{\partial t^2} + \frac{D}{b} \Delta ^2 u = 0\,
, \quad D = \frac{b^3 E}{12 (1-\sigma^2)}. 
\end{equation}
where $u$ is the displacement perpendicular to the membrane, $\rho$ is
the density of the membrane material, $E$ is the Young modulus, 
$\sigma$ is the Poisson ratio and $\Delta$ is the 2D Laplace operator. 
Searching for a solution in the form $ u \sim e^{i({\bf
    k_{\parallel}r}-\omega t)}$ 
we obtain the dispersion law for flexural waves:
\begin{equation}\label{dlawbern}
\omega_{fl}=k_{\parallel}^2 \left( D/\rho b \right)^{1/2}      \, .
\end{equation}
Substituting the conventional expressions for the Young modulus and
Poisson ratio through  
the sound velocities $c_l$ and $c_t$ and density $\rho$ we arrive at
the same dispersion law as given by Eq.~(\ref{ahlowest}). This
equation is valid only in the long wave approximation, $\lambda \gg
b$.  
\begin{figure}[t]
\begin{center}
\centerline{
\includegraphics[width=8cm]{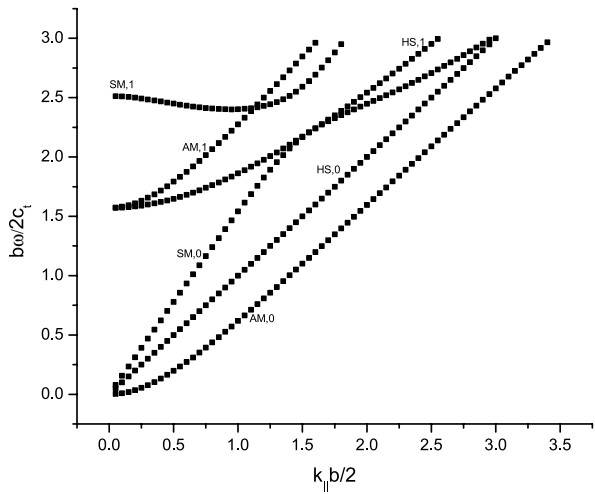}
}
\caption{The six lowest branches of vibrations in a membrane. The lowest AM branch is quadratic around zero.}
\label{branches}
\end{center}
\end{figure}

\paragraph{Heat capacity of a free standing membrane:}
We compute the heat capacity per unit area from the general equation 
\begin{equation} \label{capacity}
C = \frac {1}{A}  \sum_{s, n,\bk_{\parallel}} 
\frac{
  {(\beta \hbar \omega_{s,n (\bk_{\parallel})}})^2 e^{\beta \hbar
  \omega_{s,n} (\bk_{\parallel})}} {[e^{\beta \hbar
  \omega_{s} (\bk_{\parallel})}  - 1]^2}\, , \quad \beta \equiv
\frac{1}{k_B T}
\end{equation}
where A is the area of the membrane.

The heat capacity per unit area $C$ of a thick membrane
($b \gg \hbar \pi \beta c_t$) increases linearly with the thickness $b$  as it should for a 3D
body. But for a thin membrane  ($b < \hbar \pi \beta  c_t$) there is a
minimum of heat capacity. The position of this 
minimum depends not only on the properties of the membrane material
but also on temperature. It is important to underline that we
calculated the heat capacity of a membrane per unit area. To
find the specific heat capacity $c_v$ we should divide the heat
capacity per unit area $C$ by the thickness $b$. 

For temperature $T \ll \hbar \pi c_t/k_B b$ the heat capacity of a
membrane is dominated by the contribution of the lowest AM branch. So
we can find a crude estimate for the position of the minimum, $b_{\min}$, by 
equating the heat capacity per unit area of a bulk sample to the 
contribution of the lowest AM branch to the heat capacity per unit
area of a thin film.  

We find the total heat capacity per unit area of a bulk sample by 
multiplying the thickness $b$ of the membrane and
its specific heat capacity $c_v$,\cite{book3}
\begin{equation}\label{cap3d}
\frac{c_v}{k_B} \approx \frac {2 \pi^2}{5} \left( \frac{k_B T}{\hbar
    c} \right)^3 \, , \quad \frac{1}{c^3}\equiv \frac{1}{3} \left(
    \frac{2}{c^3_t}+\frac{1}{c_l^3} \right) \, .
\end{equation}
The contribution of the lowest AM with the dispersion law
(\ref{ahlowest}) can be easily calculated from
  Eq.~(\ref{capacity}) as
\begin{equation}\label{C0A}
\frac{C_{\text{AM},0}}{k_B}= \frac {k_B  T m^* \zeta (2)} {\pi
    \hbar^2}\, .
\end{equation}
From the equality $c_v b = C_{\text{AM},0}$ 
we obtain 
\begin{equation}\label{estmin}
b_{\min}T = a \frac{c_l \hbar^2 c^3}{ k_B^2
  c_t \sqrt{c_l^2-c_t^2}} \, , \quad a=\frac{5 \sqrt{3} \zeta (2)}{4
  \pi^3} \approx 0.11\, . 
\end{equation}
For the temperature 0.1 K the estimated position of the heat capacity
minimum $b_{\min} \approx 400$ nm.

A more accurate procedure is based on the exact expression
(\ref{capacity}). Making use of the scaling relation (\ref{eq:scal})
one can cast this equation in the form
\begin{eqnarray}
  \label{eq:cap2}
  C&=&\frac{k_B}{2\pi b^2} \cF\left(\frac{\hbar c_t}{b k_B
  T}\right)\, , \\
 \cF (z) &\equiv& \sum_{s,n}\int_0^\infty \xi \, d\xi \, \frac{[z
 w_{s,n}(\xi)]^2} {\sinh^2 [z
w_{s,n}(\xi) ]} \, .\label{eq:cap2a}
\end{eqnarray}
Here $\xi \equiv b k_\parallel$ and we assume that the temperature is
well below the Debye temperature so that we can 
integrate from 0 to infinity. One
sees that the heat capacity per unit area is a nonmonotonous 
function of the film
thickness, $b$. Its minimum can be determined by equating the
derivative $\partial C/\partial b$ to zero. It leads to the relation 
\begin{equation}
  \label{eq:opt1}
  b_{\min}\cdot T = \frac{\hbar c_t}{k_B z^*}
\end{equation}
where  $z^*$ is determined from the equation
$\cF'(z^*)z^*=-2\cF (z^*)$. The value of 
$z^*$ depends on the ration $c_t/c_l$ and for $c_t/c_l =1.6$, $z^*
\approx 1.9$.

\begin{figure}[t]
\begin{center}
\includegraphics[width=8cm]{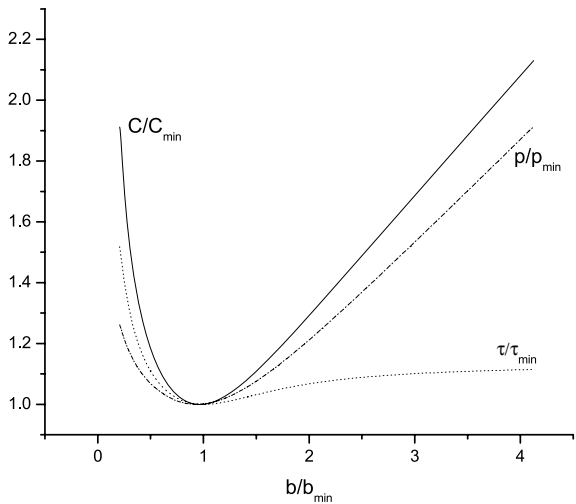}
\caption{The heat capacity per unit area $C$ of a membrane, the response time $\tau$ and the derivative of the radiation power with temperature $p$ as a function of 
its thickness $b$. $b_{\min}$ is the thickness for which the heat capacity reaches its minimal value.  We used the values  $c_t = 6000$~m/s, $c_l /
c_t=1.6$, which are  
approximately the same as in the experiments\protect{\cite{paper2}} 
on SiN$_x$ membranes.  }
\label{capacity2}
\end{center}
\end{figure}

Numerical calculations including the 30 lowest branches give the value
$b_{\min} =240$ for $T=0.1K$, which is not far from the crude estimate obtained
  using only the lowest mode. Substituting (\ref{eq:opt1}) into
  Eq.~(\ref{eq:cap2}) we obtain
  \begin{equation}
    \label{eq:cap3}
    C_{\min}=\frac{(k_BT z^*)^2}{2\pi (\hbar c_t)^2} k_B \cF (z^*)\, .
  \end{equation}
The result can be summarized in the universal form  
\begin{equation}
   \frac{C}{C_{\min}}=\left(\frac{b_{\min}}{b}\right)^2 \, \frac{\cF(z^*
  b_{\min}/b)}{\cF(z^*)}
\end{equation}
where $b_{\min} (T)$ is given by Eq.~(\ref{eq:opt1}). This is shown in
Fig.~\ref{capacity2}. 

\paragraph{Applicability range of the theory:}

There are several phenomena limiting the increase of the heat capacity
with decrease of the film thickness. The obvious limit is that the
elasticity theory is not applicable when the thickness is comparable
with interatomic distance. 
Below we will discuss another limit posed by anharmonic effects. 
The amplitude of thermal vibrations increases with decreasing thickness.
As a result,  in a very thin membrane the
harmonic approximation used in this paper becomes invalid. 

To make an estimate, let us
assume for simplicity that thickness of 
the membrane $b \ll b_{\min}$ and take into account only the  AM,0 branch. 
Due to the thermal vibrations, the membrane will not be perfectly flat 
but have a shape described by some function  $f(x,y)$. The resulting 
increase in the area of the membrane corresponds to some average 
tension, $\cT$, of the membrane. Let us therefore first study the 
membrane in the presence of an external tension, and later insert the 
thermal value of this tension. 
In this case in the equation of motion (\ref{landau}) an additional term
appears:
\begin{equation}\label{tension}
\rho \frac{\partial ^2 u}{\partial t^2} = -\frac{E b^2}{12(1-\sigma^2)}
\frac{\partial ^4 u}{\partial x^4}+\cT \frac{\partial ^2 u}{\partial
 x^2}\, ,
\end{equation}
the dispersion law being
\begin{equation}\label{ndisplaw}
\omega = \sqrt{\frac{E b^2}{12 \rho (1-\sigma^2)}k^4 + \frac
  {\cT}{\rho}k^2}\, . 
\end{equation}
When the second term in (\ref{tension}) becomes greater than the
first one the dispersion law changes to linear and the heat capacity of
the membrane approaches some limiting value. 
Since thermal vibration produce an average strain they will
change the dispersion law. So for a crude estimate we will evaluate
the average strain produced by the thermal vibrations and compare
its contribution with the first item $ E b^2 k^4/12 \rho(1-\sigma^2)$.
We can express variation of the membrane area as 
$\Delta A \approx \frac{1}{2}\int_A |\nabla f|^2 \, dxdy$.
Substituting for $f$ thermal modes of a square membrane and performing
thermal average one obtains an estimate for the typical elongation:
\begin{equation}
  \label{eq:elong1}
  \frac{\Delta L}{L}  \approx \frac {k_B T}{\rho
  b^3(c_l^2-c_t^2) }\, \frac{ c_l^2}{c_t^2 }\ln\frac{L}{L_T}\, , \quad 
L_T = \sqrt{\frac{2\pi ^2 \hbar^2}{m^* k_B T}}\, .
\end{equation}
This relative elongation strongly depends on the membrane
thickness. 
Substituting $\cT \approx E (\Delta L/L)$ 
in Eq.~(\ref{tension}) 
and comparing its contribution with the first item for thermal wave
vectors we obtain an estimate of the critical thickness below which
out theory is not applicable:
\begin{equation}\label{limit}
b_{\text{cr}} = \left(\frac{12 \hbar c_l}{c_t \sqrt{3(c_l^2-c_t^2)}
\rho}\cdot\ln\frac{L}{L_T}\right)^{1/4}\, .
\end{equation}
Numerically, this estimate gives the thickness of several
atomic layers, so it does no pose additional limitations comparing to
general range of applicability of the elasticity theory.

\paragraph{Sensitivity and response time of the bolometer:}
\begin{figure}[t]
\begin{center}
\includegraphics[width=4.5cm]{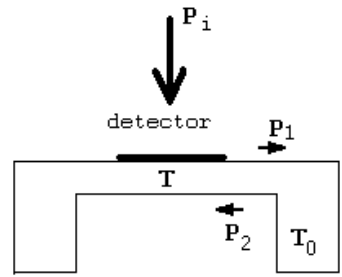}
\caption{Schematic of a free membrane with mounted detector. 
\label{fig3}}
\end{center}
\end{figure}
Suppose we use a thin free standing membrane with a detector on it as
a sensitive element of  a bolometer. 
In presence of incident flow of energy,
$P_i$, see Fig.~\ref{fig3}, we can obtain the
temperature of a membrane as 
a function of time:
\begin{equation}\label{time}
P_i(t)= P_1(T)-P_2(T_0)+C_{\text{mem}}(T)(dT/dt)
\end{equation}
where $C_{\text{mem}}$ is the heat capacity of the membrane under the detector, 
$P_2(T_0)$ is the heat flow from surrounding media to the membrane under the detector, 
$P_1(T)$ is the heat flow from the membrane under the detector and
$T_0$ is the temperature 
of surrounding media.We assume that phonons are ballistic,\cite{paper3}
 i. e., the mean free path of the phonons is much longer than the size
of the detector.  
Usually the relative temperature difference is less than 0.01, so we can expand the radiation power $P_1(T)$ around $T_0$ and use heat capacity as a constant.
In this case we obtain a response time $\tau = C/(\partial P/\partial T)$.   
In equation (\ref{eq:cap2}) the part that depends only on $bT$ can
be separated 
\begin{equation}\label{transc}
C=T^2\tilde{C}(bT)
\end{equation}
The equation for power radiation \cite{paper3} can be similarly transformed to
\begin{equation}\label{transp}
P=T^3\tilde{P}(bT)\, .
\end{equation}
{}From this we can obtain response time is a function only of $bT$
\begin{equation}\label{resptime}
\tau = \frac{\tilde{C}(bT)}{3\tilde{P}(bT)+bT\tilde{P}'(bT)}\, .
\end{equation}

From this we immediately obtain that the minimal response time,  $\tau_{\min}$,  as a function of membrane thickness does not depend on temperature.
The dependence of the response time $\tau$  versus the membrane
thickness is shown on the Fig.~\ref{capacity2}. The position of
$\tau_{\min}$  
is close to the position of $C_{\min}$. On Fig.~\ref{capacity2} this
difference is difficult to see because it is only 5\%. 
If we measure pulses of energy shorter than $\tau$ the rate of
temperature increase depends only on the heat capacity of the
membrane.  
In the opposite case when pulses are longer than $\tau$ the final temperature difference depends only on $p=(\partial P/\partial T)$.
The dependence of $p$ on the ratio $b/b_{\min}$ is also presented on Fig.~\ref{capacity2}. 
From this picture we can see that the heat capacity $C$, the derivative of the radiation power with temperature $p$ and response time $\tau$
reach their minimal values at different thicknesses. This difference is about  10\%. 
If one want to construct a detector he wants to obtain either the shortest response time or the greatest temperature difference.
Our calculations demonstrate that the best sensitivity and the shortest response time can be obtained by choosing the supporting membrane with 
thickness $b_{\min}$.

\paragraph{Discussion and conclusions:}

We have studied the heat capacity of a thin membrane at low temperatures
such that the typical wavelength of thermal phonons is of the order or 
smaller than the thickness of the membrane. Because of the quadratic 
dispersion law of the lowest vibrational branch, the heat capacity 
per area will increase with decreasing thickness below a certain 
thickness $b_{\min}$. The thickness of minimal heat capacity is temperature 
dependent, $b_{\min}\propto1/T$. 
The shape of the curve $C(b)$ has the universal form  (\ref{eq:cap3}). 
If we want to use the membrane for support of microbolometers, the 
reduction of the heat capacity is important for sensitivity of the
detector for short pulses,
 and the thickness should be chosen equal to $b_{\min}$ 
at the operating temperature of the bolometer. 

\acknowledgments
This work was partly supported by the Norwegian Research Council
(through NANOMAT and STORFORSK programs) and by the U. S. Department of
Energy Office of Science through contract No. DE-AC02-06CH11357.

\end{document}